\documentclass[11pt,twoside]{article}
\usepackage{asp2004}
\usepackage{psfig}
\usepackage{epsf}
\def\edcomment#1{\iffalse\marginpar{\raggedright\sl#1\/}\else\relax\fi}
\reversemarginpar

\def\nifsx{${}^{56}$Ni}
\def\msun{M_\odot}

\def\Msun{$\msun$}
\def\rsun{R_\odot}
\def\Rsun{$R_\odot$}

\def\rfig#1{Fig.\ref{#1}}

\begin{document}
\title{Light Curve Models of Supernovae and X-ray spectra of Supernova
 Remnants}
\author{S.I.Blinnikov, P.V.Baklanov, A.V.Kozyreva, E.I.Sorokina}
\affil{ITEP \& SAI, Moscow, Russia, and  MPA, Garching, Germany}

\begin{abstract}
We compare parameters of well-observed type II SN1999em derived by M.Hamuy
and D.Nadyozhin based on Litvinova \& Nadyozhin (1985) analytic fits with those found
from our simulations using our radiative hydro code {\sc stella}. 
The same code applied to models of SN1993J allows us
to estimate systematic errors of extracting foreground extinction toward
SN1993J suggested by Clocchiatti et al.~(1995) which is based on the
assumption of black body radiation of the supernova envelope near the first
maximum light after shock break out.
A new implicit two-temperature  hydro code code {\sc supremna} is introduced which
self-consistently takes into account the kinetics of ionization, electron
thermal conduction, and radiative losses.
Finally, a combination of  {\sc stella} and {\sc supremna} allows us to use the same SNIa models both for building their light curves and predicting X-ray
spectra of young Supernova remnants such as Tycho and Kepler. 
For the comparison of theoretical results with the observations we used data on
Tycho SNR obtained with XMM--Newton space telescope.
\end{abstract}
\thispagestyle{plain}

\section{Introduction}

Three topics will be discussed briefly here.
First, we show a way for extracting SN~II parameters from \ubv\ light curves (using SN1999em as an
example). Then we turn to estimating foreground extinction  (example of SN1993J).
And finally, we demonstrate using SN~Ia models for predictions both of  \ubvri\ light curves of SNe and  X-ray spectra  of young SNRs. 

\section{SN~II light curve theory {\it vs.} observations}

Nadyozhin~(2003) and Hamuy~(2003) have obtained masses ${\cal M}$, radii $R$, and explosion energies $E$ for a set of SNe~II. 
Both have taken the observed values of $M_V, \Delta t, u_{\rm ph}$ on the light curve plateau and extracted  ${\cal M}, R, E$, based on the relations found by Litvinova \& Nadyozhin (1983, 1985), --- LN85 hereafter.

Let us look into parameters found by Nadyozhin~(2003) and Hamuy~(2003) for
an example of well-studied SN~1999em.
Note that Nadyozhin's~(2003) Eq.(1) changed units of $E$  from LN85
and 2 typos are introduced by MN into this paper; see the original version \texttt{astro-ph/0303411}.

\begin{table}[!ht]
\smallskip
\begin{center}
{\small
\begin{tabular}{llll}
\tableline 
\noalign{\smallskip}
 Input & Nadyozhin~(2003) & Hamuy~(2003) \\
\noalign{\smallskip}
\tableline
\noalign{\smallskip}
$D$      &  12.38   &        10.7 &       Mpc \\
$M_V$  & -16.78   &        -16.44 &    \\      
$\Delta t$ & 110  &    124   &  days \\
$u_{\rm ph}$ & 2900   &   3290  &  km/s \\
\noalign{\smallskip}
\tableline
\tableline 
\noalign{\smallskip}
Derived &\phantom{Nadyozhin~(2003)} & \phantom{Hamuy~(2003)}\\
$E$ & 0.63 &  $1.2^{+0.6}_{-0.3}$ & foe \\
${\cal M}$ &   13.2 & $27^{+14}_{-8}$ &  \Msun \\
$R$ & 569  & $249^{+243}_{-150}$ & \Rsun \\
\noalign{\smallskip}
\tableline
\end{tabular}
}
\end{center}
\end{table}

From the Table 
we see that the discrepancy of derived parameters
for the same supernova is large.
While input parameters differ within 10\% only, the derived ones
may be in disagreement by a factor of 2. 
There is an appreciable difference in the assumed distance 
$D$ in the two papers, and hence in $M_V$.
The values of $E$, ${\cal M}$, and $R$ 
 scale with the distance as:
\begin{equation}
  E\sim D^{-0.675},\quad {\cal M}\sim D^{-1.17},\quad R\sim D^{2.86}\, .  
\end{equation}
 Thus, it is very important to know $D$ with as high accuracy as 
 possible.  
But the difference may be explained by different $D$ only for 
$R$, but not for $E$ and ${\cal M}$: other small differences are also
important.
Hence, the LN85 process is rather unstable.

If one looks into the \ubv\ light curves in 
Litvinova \& Nadyozhin (1983) one can see that they are all similar with
a pronounced plateau, 
because they were built in a simple equilibrium diffusion approximation,
while well observed SNe~II have a good plateau
only in {\it V} and redder filters, but not in 
{\it B} and {\it U} -- see, e.g. 
Leonard et al.~(2002) and \rfig{HNLCs} below.

This means that more detailed models are needed in order to
reproduce the observed \ubv\ behavior.
We have considered an example of the well-studied SN~1999em with a goal
to extract its parameters using more sophisticated modern multi-group SN~II LC modeling.
We have used our code {\sc stella} (Blinnikov et al. 1998; Sorokina \& Blinnikov 2003, 2004). See Baklanov (2002)
where a theoretical catalog of \ubv,
$M_{\rm bol}$ light curves is built for very different SN~II presupernovae
for a set of ${\cal M}, R, E$. 
In addition to those parameters used by LN85 the mass ${\cal M}_{\rm Ni}$
of radioactive \nifsx\  is also accounted for.

To model SN~1999em
we construct non-evolutionary preSN models with extended hydrogen-rich envelopes, similar to LN85, but a bit more
realistic because they have a compact massive core and \nifsx\ mixed to the
envelope as in the models for SN~1987A (Blinnikov et al. 2000). 
The envelope has mass fraction of H,  X=0.7, and the fraction of metals was assumed
Z=0.03 for the first models (in proportion to cosmic distribution with this Z). 

The \rfig{HNLCs} shows our model light curves when the presupernova parameters
(and reddening) are those
of Hamuy~(2003) \& Nadyozhin~(2003).
We see that fluxes on the plateau disagree up to two stellar magnitudes 
with observations.  
\begin{figure}[!ht]%
\plottwo{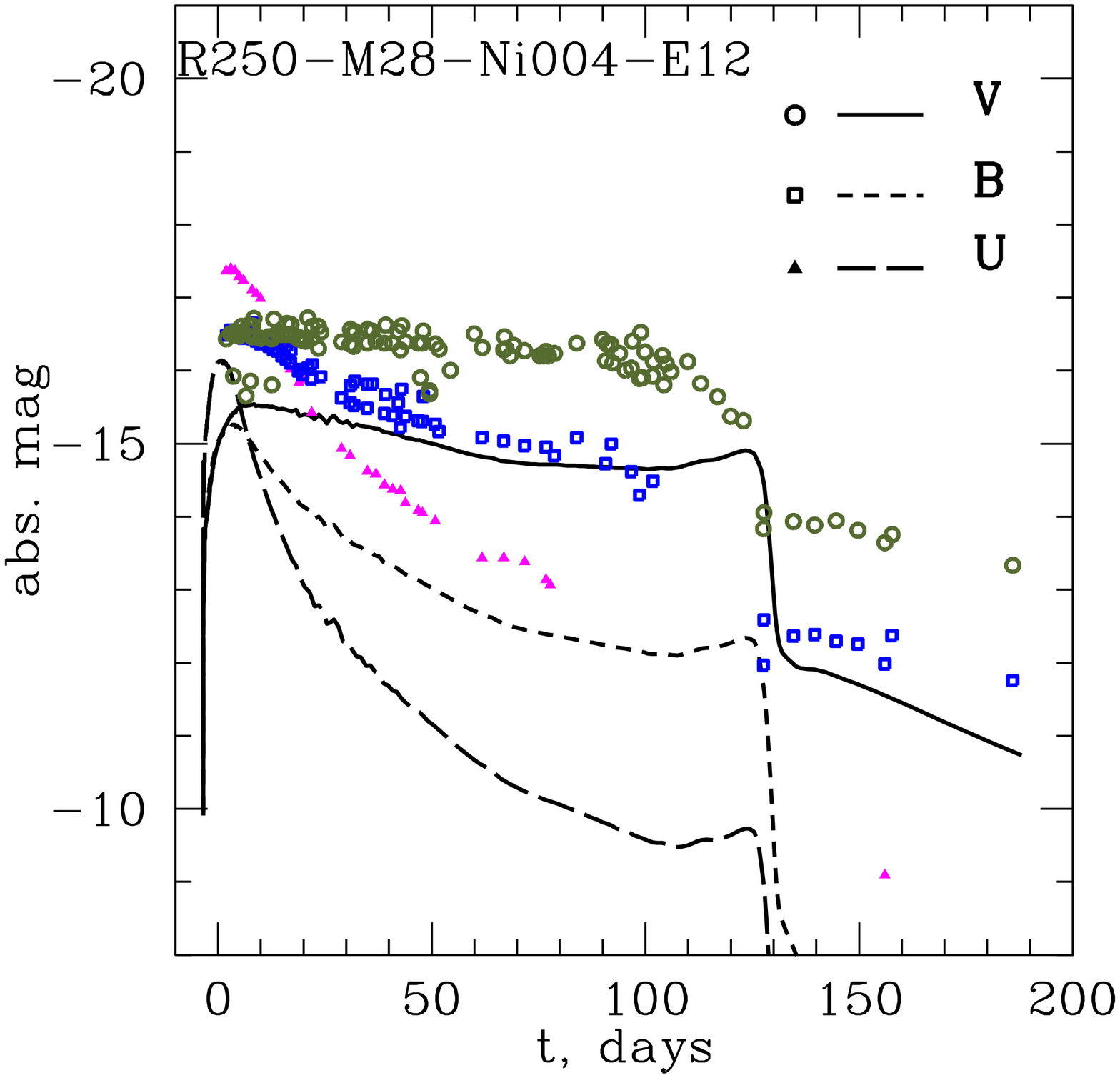}{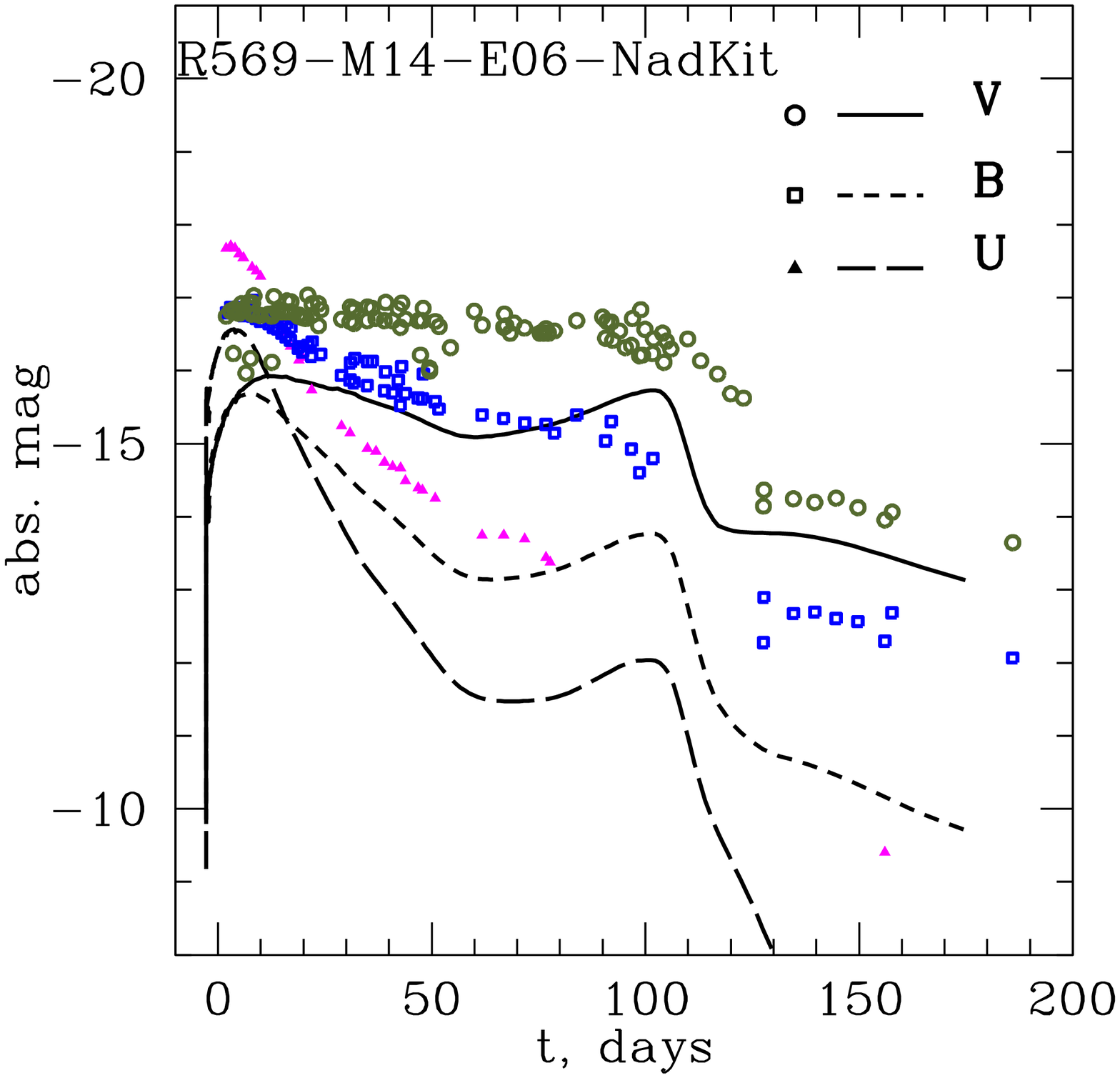}%
\caption{\ubv\ for SN~1999em parameters = Hamuy (left), Nadyozhin (right)}%
\label{HNLCs}
\end{figure}

\begin{figure}[!ht]%
\plottwo{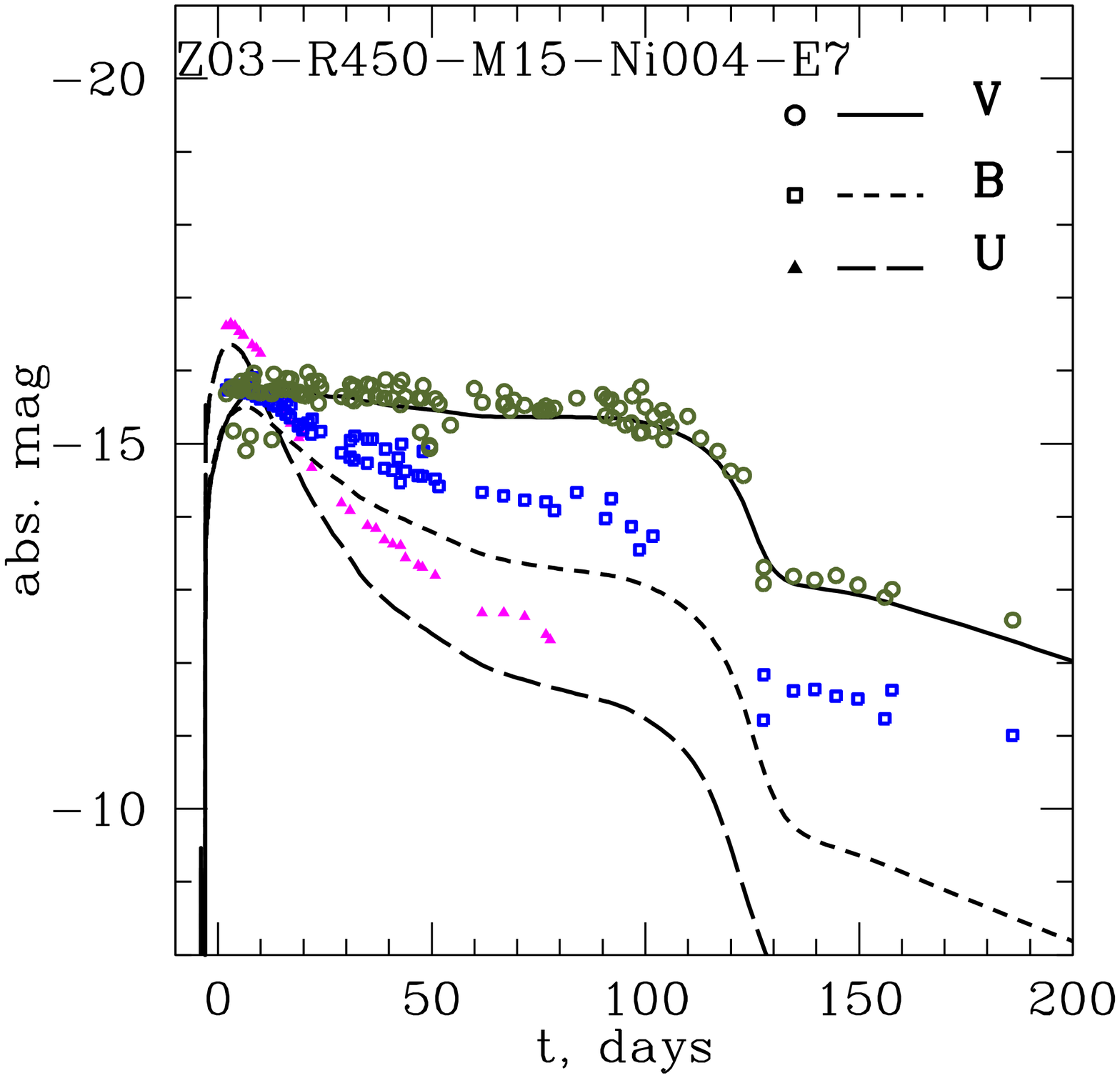}{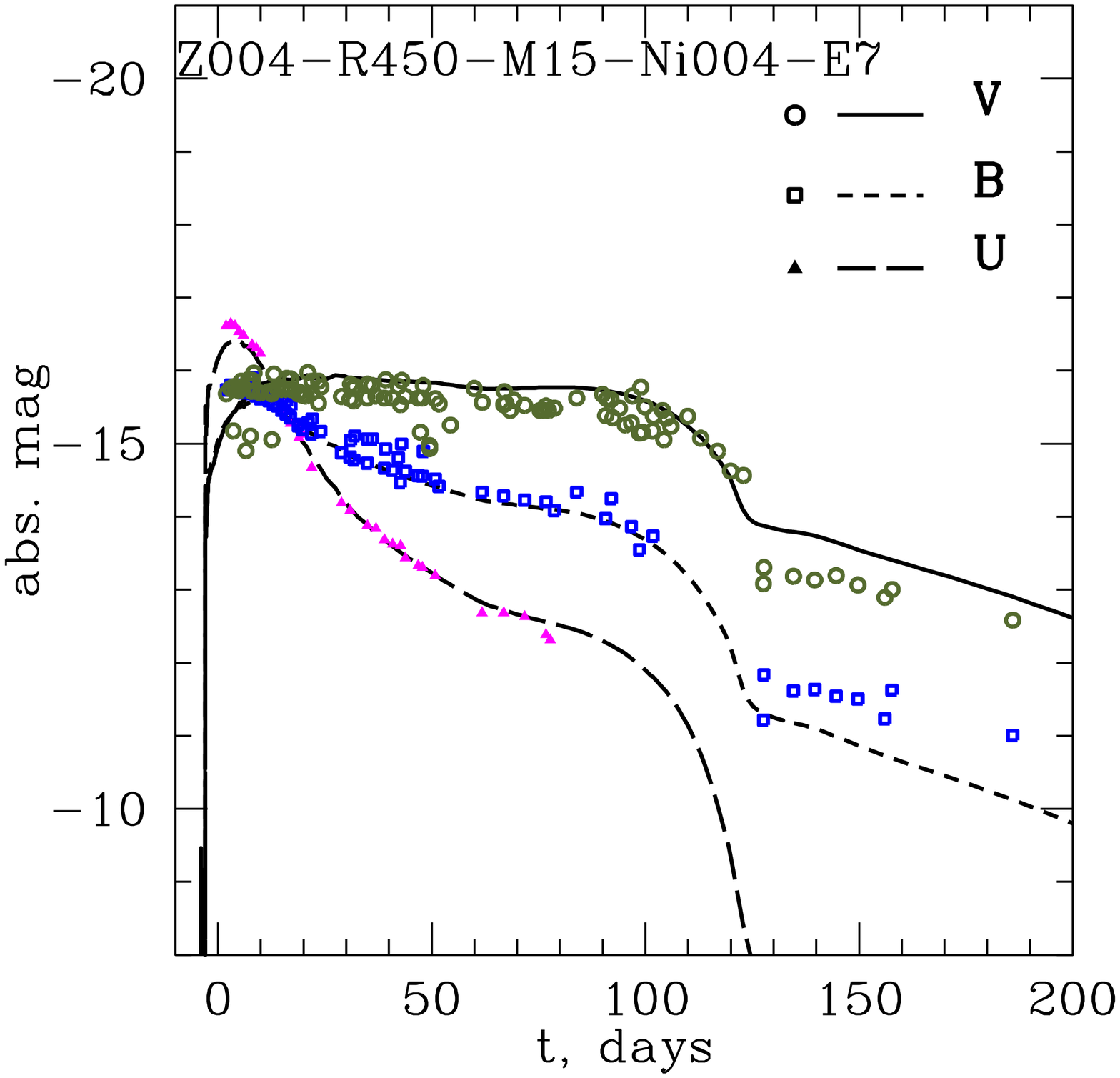}%
\caption{\ubv\ short  distance scale:
$D= 7.5$~Mpc, metals Z=0.03 (left), Z=0.004 (right). }%
\label{highlowZ}
\end{figure}

In search of best-fitting model were have varied the input parameters
in wide range and found a reasonably good fit for
 ${\cal M}=15 \msun, \; R=450 \rsun, \; E=7\times 10^{50} $ ergs,
 ${\cal M}_{\rm Ni}=0.04\msun$ (\rfig{highlowZ}, left).
This set is good only for $D=7.5$~Mpc (EPM, Hamuy et al. 2001),
which is much shorter scale than the Cepheid distance found by 
Leonard et al. (2003).
A very nice fit is found for the same parameters, but for low
metal abundance, Z=0.004 (\rfig{highlowZ}, right).
Not only \ubv, but also the dependence of photospheric speed
$ u_{ph}(t)$ is good for the last model.
Low metal abundance agrees well with 
the results of spectral modeling by Baron et al.~(2000).
For the long distance scale  one can get a satisfactory \ubv\  only for huge $R\sim 10^3 R_{\sun}$. Work on this is in progress.

\section{Extinction toward SN~1993J}

\begin{figure}[!ht]%
\plottwo{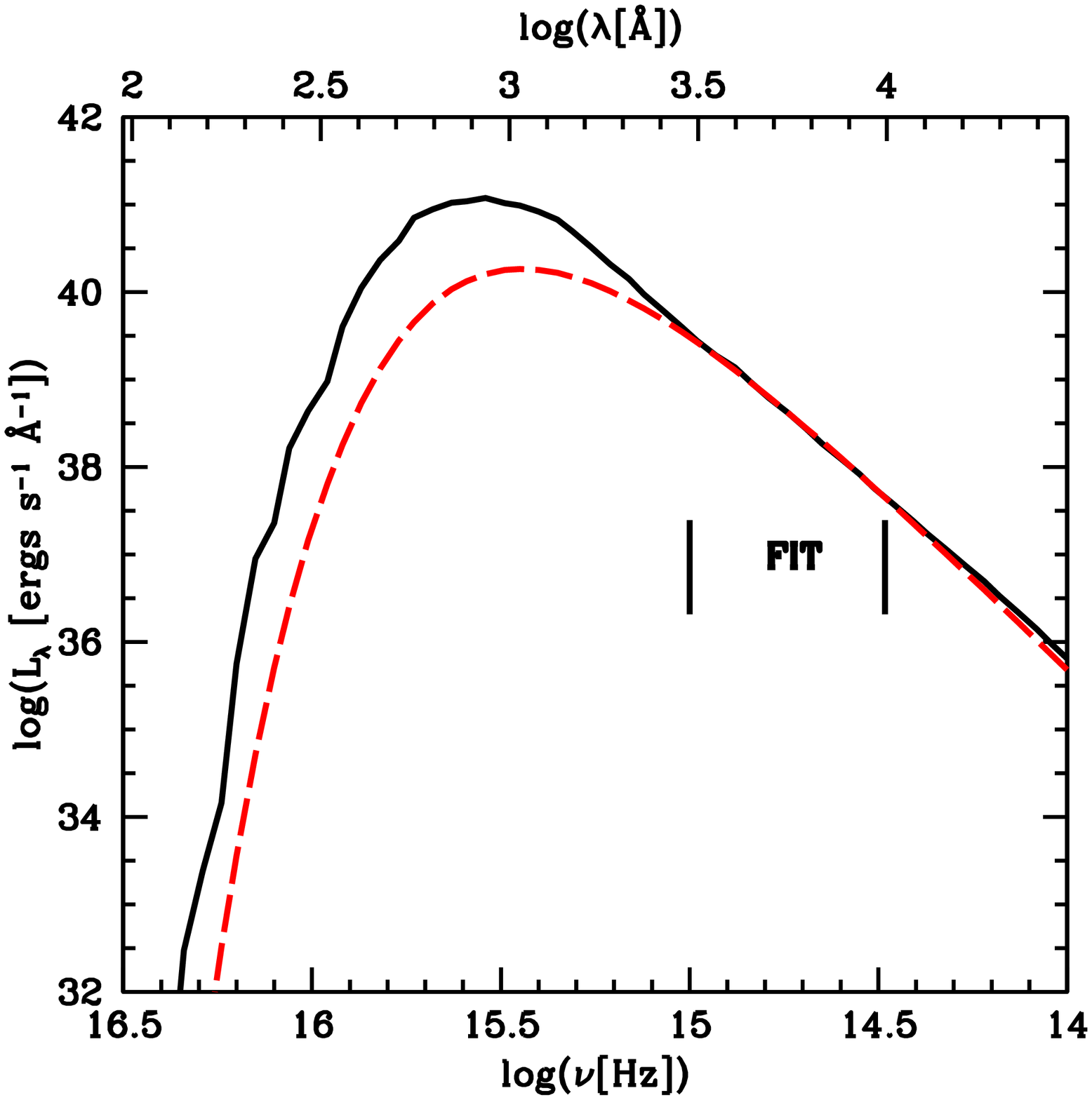}{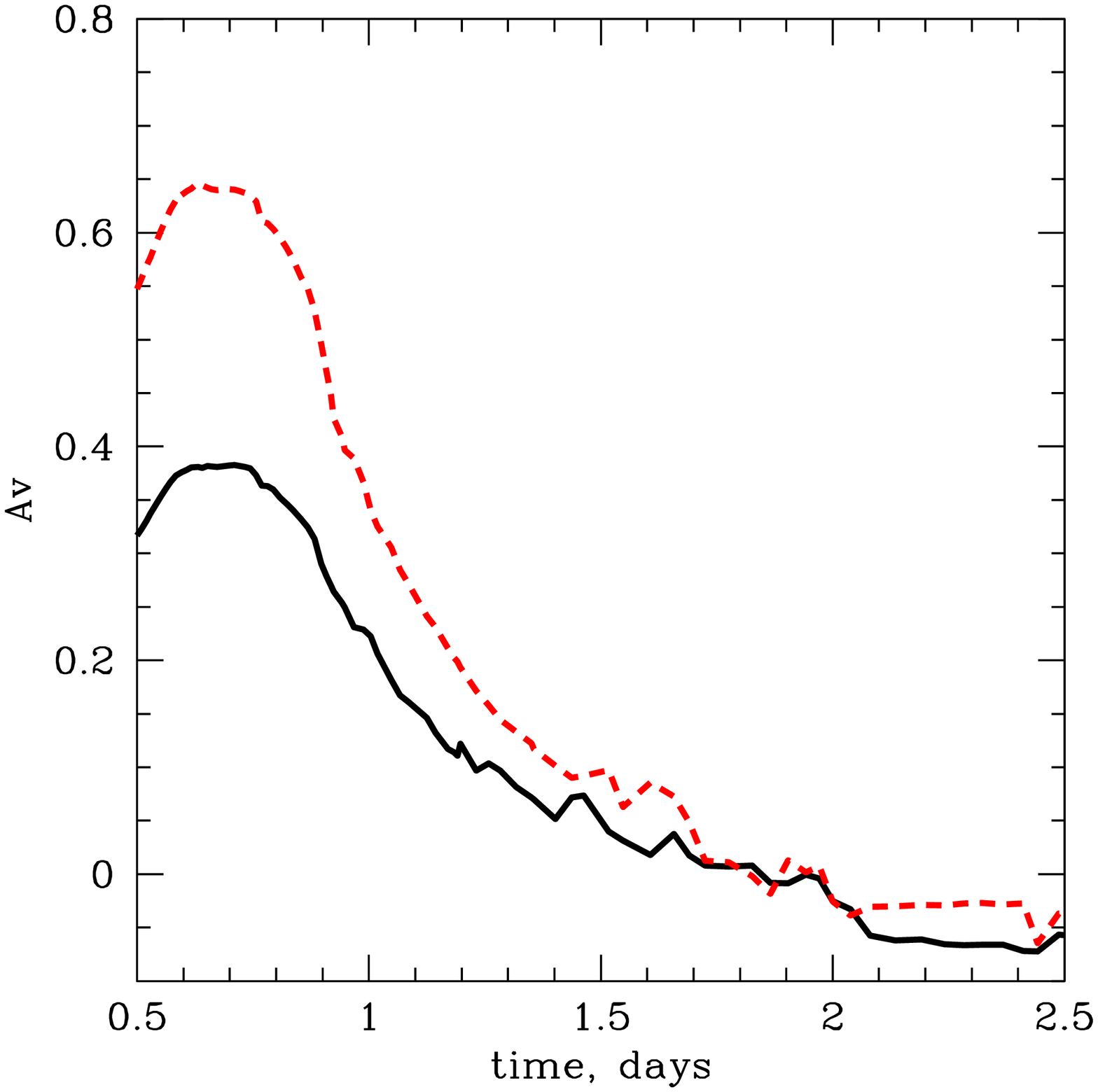}%
\caption{Left. 1.12 days after SN explosion. Solid: model. 
Dashed: black body fitted to the model
in the visible range (labeled `FIT') with 
Cardelli et al. (1989) absorption law. The best fit gives $T=39300$~K
and $A_V=0.241$ (while true absorption in this numerical experiment is zero!).
Right. Dependence of faked extinction parameter $A_V$ on time for
different extinction laws: solid for a simple exponential law and dashed for
Cardelli et al. (1989)
}%
\label{bbav}
\end{figure}

The determination of foreground extinction toward supernovae (SNe) is very
important.
A technique for estimating the extinction has been suggested
by Clocchiatti et al. (1995) which is based on the
assumption of black body SN radiation near the
first maximum light after shock breakout. 
Here we 
note that the true spectrum
of an SN after the shock breakout, theoretically, is not a pure black body
(due to importance of scattering and non-gray absorption in the ejecta).
Using spectra which were calculated by the {\sc stella} code
for the 13C model of SN~1993J (Woosley et al. 1994; Blinnikov et al. 1998)
and which fit energy distribution in the
observer frame sufficiently well, we found that the values of
$A_V$ may differ substantially from found under the black body assumption
(\rfig{bbav}).

\section{SN~Ia models: light curves and Tycho Supernova Remnant}

\begin{figure}[!ht]%
\plottwo{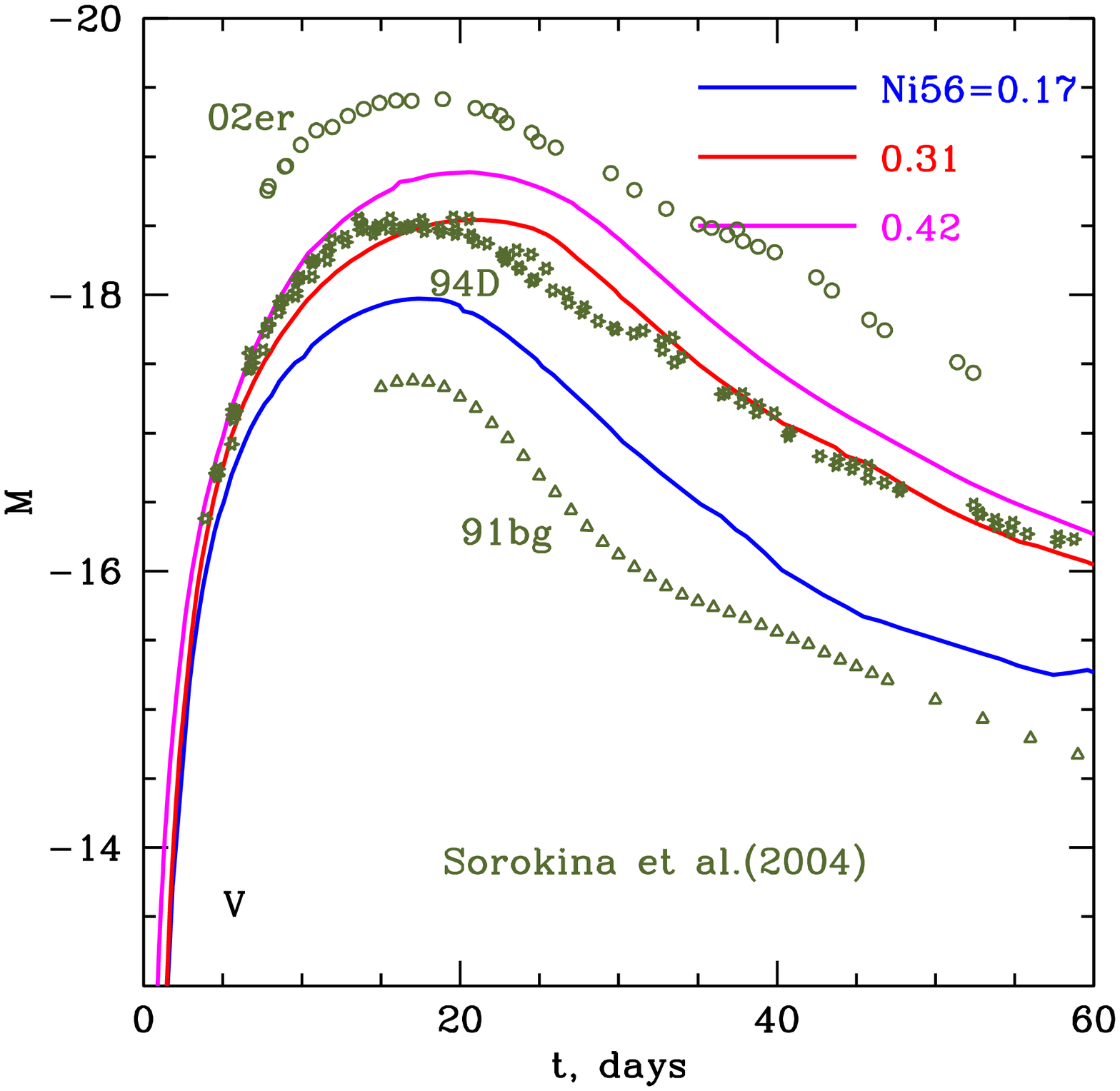}{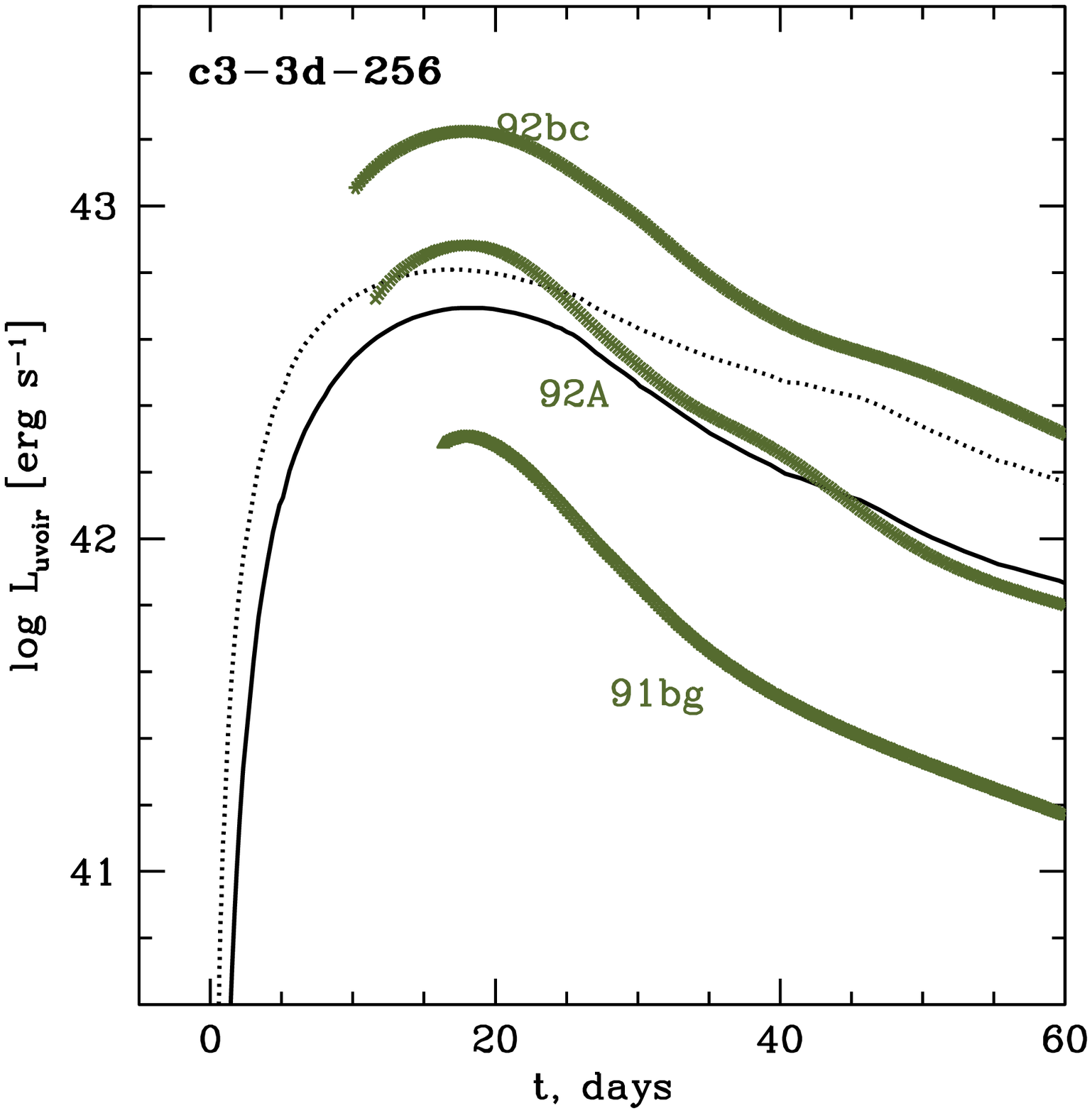}%
\caption{Left. {\it V} light curves for new MPA models.
Right. Bolometric light curves for c3\_3d\_256
(dotted $L_{\rm bol}$, solid $L_{\rm UVOIR}$) }
\label{lcc3}
\end{figure}

Here we concentrate on 3D SN~Ia models computed at MPA (Reinecke, Hillebrandt, \& Niemeyer 2002). 
Our assumptions on their light curve models using {\sc stella} are described
in Sorokina \& Blinnikov (2003), Blinnikov \& Sorokina (2004).
The composition of early MPA models was rather crude.
Recently Travaglio et al. (2004) computed
a detailed explosive nucleosynthesis for some of them.
We show {\it V} light curves for those models 
and a bolometric flux for centrally ignited c3\_3d\_256 in \rfig{lcc3}. 
The curves are in reasonably good agreement with observations.

It is interesting to compare the properties of young SNRs, such as Tycho,
with predictions of the same models that have been used for LC modeling.
For the first hundreds years the pattern of flow has a forward shock sweeping  up
ISM material and  a reverse shock propagating into the slower-moving ejecta
which we call Nadyozhin-Chevalier stage:
a self-similar solution by Nadyozhin (1981, 1985) and Chevalier (1982) is much more realistic than widely used Sedov solution.
For this stage  Sorokina et al. (2004) have 
developed a new 1D Newtonian implicit hydro code {\sc supremna}  which allows 
us to study 
dynamics and radiation of young SNRs without using approximate solutions.

The main features of the code:
 ionization energy is taken into account in the equation of state;
 electron thermal conduction with saturation at the sound speed
and radiative losses are not neglected;
 $T_e \ne T_i$ (equilibration by Coulomb collisions).
The X-ray spectrum
is computed by a time-dependent ionization code
which calculates evolution of the ionization stages for every mesh zone 
at \emph{each} hydro time step for all ions of 15 elements; it
takes into account collisional ionization, autoionization, 
photorecombination, dielectronic recombination, and charge transfer. Comparison with other codes used for young SNRs is given in the table.


\begin{tabular}{l|c|c|c|c|c}

{\footnotesize Work}               &Hyd--        & Ioni-- & Rad.& Th. & $T_e$   \\                                     &ro      & zation & loss & cond. &  vs. $T_i$ \\
\hline
{\footnotesize Hamilton \& Sarazin(1984)} & 0 & Neq        & +          & --       & 2--3$T$ \\
\hline
{\footnotesize Itoh et al.(1988)}       & 1  & Neq        & --         & -- (+)   & 2$T$ \\
\hline 
{\footnotesize Brinkmann et al.(1989)}  & $\to$  & Neq        & --         & --       & 1$T$ \\
\hline 
{\footnotesize Badenes et al.(2003)}   & $\gamma$  & Neq        & --         & --       &  2$T$ \\
\hline 
{\footnotesize Sorokina et al.(2004)}    & $\to$   & Neq        & +          & +        & 2$T$ \\
\hline
\end{tabular}

\begin{figure}[!ht]%
\plottwo{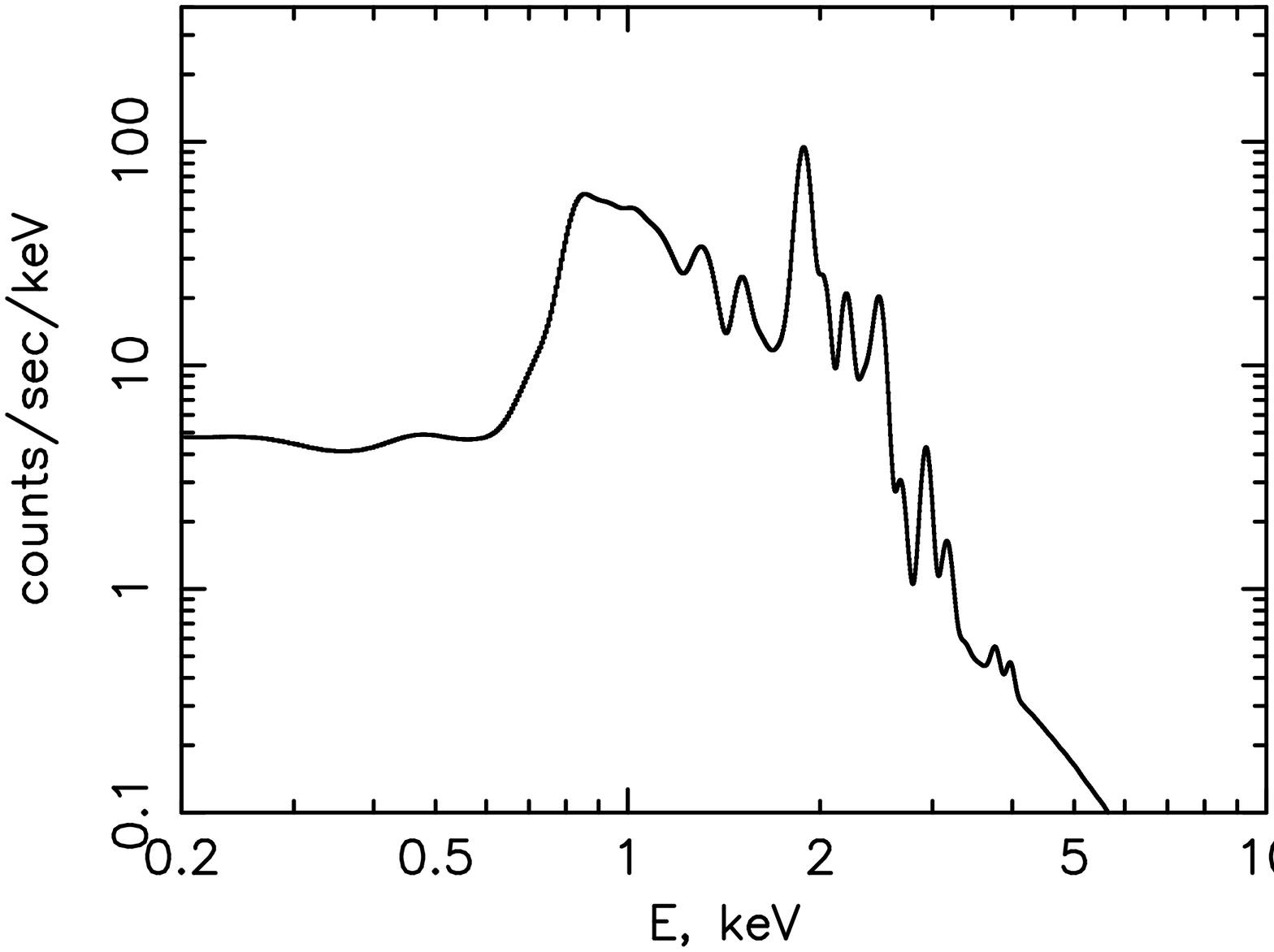}{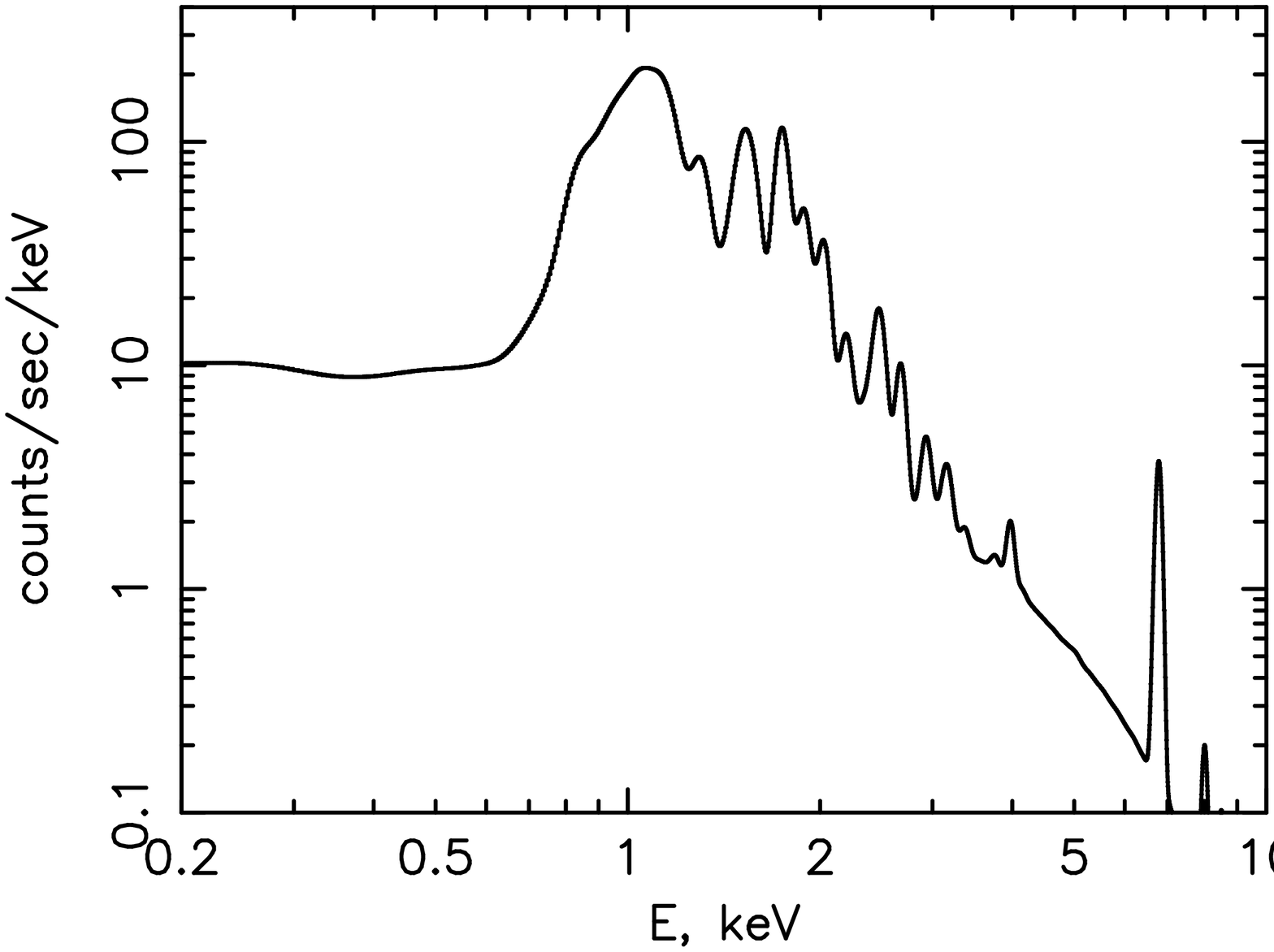}
\caption{X-ray spectrum for W7, convolved with the response matrix
of  EPIC PN on XMM--Newton (left), the same for MPA model MR0
(right). 
}%
\label{spw7mr0}
\end{figure}

As an example, we show in \rfig{spw7mr0} the X-ray spectra for the age
of Tycho 430 years for W7 model and for
the same MPA model MR0 and for the same parameters of ISM as used by
Blinnikov \& Sorokina (2004) and Kosenko et al. (2004). 
The 1D W7 model has a problem reproducing iron K$_\alpha$ emission
without an additional mixing, as noted already by Itoh et al. (1988),
while MPA models are well mixed from the beginning and they show
a prominent iron line. There may be a problem with lines of
intermediate mass metals, such as Ca and Si, but the development
of thermonuclear 3D deflagration models may resolve this problem.

An important result of our work is the development of thermal instability
in metal-rich ejecta in the models with low electron conduction.
This confirms the old result found by Hamilton \& Sarazin (1984) which
was ignored in numerical models of young SNRs.

\section{Conclusions}

\begin{itemize}
\item To get SN parameters, one \emph{has} to craft a
full light curve model and $v_{\rm ph}(t)$. 

\item In case of SN~1999em we easily find reasonable parameters and low
$Z$ for EPM distance of 7.5 Mpc. The Cepheid distance requires a larger
$R$ of the presupernova.

\item Using black body assumption for early SN spectra may introduce a systematic error into estimates of $A_V$.

\item New MPA 3D models of SN~Ia 
produce faster bolometric LCs, closer to observed ones.

\item Current 3D deflagration MPA models can easily explain iron in spectra
of young SNRs.  

\item Energy losses may be \emph{very} important in the evolution 
of metal-rich ejecta of SN~Ia.  
They must be taken into  account in realistic simulations of their dynamics.

\end{itemize}

\acknowledgments
The work is supported by grants RFBR 02-02-16500a, 04-02-16793a,
NSF  AST-02--06111, NASA  NAG5-12D36, and by MPA (Garching) visitor program. 
SB and ES are grateful to W.~Hillebrandt and
to S.~Woosley for their hospitality.
SB thanks F.R\"opke for help on density structure of MPA models and organizers of this meeting in Padua and the INT
at the University of Washington (program INT-04-2) for their support.

\end{document}